\documentclass[%
 reprint,
 preprintnumbers,
 amsmath,amssymb,
 aps,
]{revtex4-2}
\pdfoutput=1
\usepackage[pdftex]{graphicx}% Include figure files
\usepackage{dcolumn}% Align table columns on decimal point
\usepackage{bm}% bold math
\usepackage{braket}
\usepackage{cancel}
\usepackage{bm}
\usepackage{amsmath, amssymb}
\usepackage{here}
\usepackage{qcircuit}
\usepackage{float}
\usepackage{color}
\usepackage{url}
\usepackage{hyperref}% add hypertext capabilities
\begin{document}

\preprint{OU-HET-1091}

\title{Quantization of Blackjack: \\ Quantum Basic Strategy and Advantage}

\author{Yushi Mura}\email{y\_mura@het.phys.sci.osaka-u.ac.jp}
\affiliation{Osaka University, Toyonaka, Osaka 560-0043, Japan}

\author{Hiroki Wada}\email{hwada@het.phys.sci.osaka-u.ac.jp}
\affiliation{Osaka University, Toyonaka, Osaka 560-0043, Japan}


\begin{abstract}
Quantum computers that process information by harnessing the remarkable power of quantum mechanics are increasingly being put to practical use. In the future, their impact will be felt in numerous fields, including in online casino games. This is one of the reasons why quantum gambling theory has garnered considerable attention. Studies have shown that the quantum gambling theory often yields nontrivial consequences that classical theory cannot interpret. We devised a quantum circuit reproducing classical blackjack and found possible quantum entanglement between strategies. This circuit can be realized in the near future when quantum computers are commonplace. Furthermore, we showed that the player's expectation increases compared to the classical game using {\it quantum basic strategy}, which is a quantum version of the popular basic strategy of blackjack.
\end{abstract}

\maketitle
\onecolumngrid

\section{\label{intro}Introduction}
Quantum information science is growing rapidly not only in theory but also at the hardware level; this leads us to believe that quantum computers will soon be practically applied on a large scale. Once quantum computers are realized on a commercial scale, quantum technology (such as devices or communications) will be widespread. In such a future, an interesting question that arises is how quantum computers can change the world. For example, in quantum game theory, Meyer \cite{PhysRevLett.82.1052} showed that for a two-player zero-sum strategic game, a player's expectation can increase when using quantum strategies. After Meyer's work, Eisert, Wilkens, and Lewenstein (EWL) \cite{PhysRevLett.83.3077} showed that quantum entanglement changes the Nash equilibrium \cite{Nash48} between the strategies of two persons in Prisoners' Dilemma (PD). The strategy of whether to cooperate or defect adopted by one person affects another person's strategy through quantum entanglement between the two individual strategies. Prisoners can choose quantum strategies, which are described by unitary operators; therefore, they have more strategies at hand compared to the classical game and can use them to increase their profit. There are many studies inspired by their quantization scheme (e.g. summarized in \cite{doi:10.1142/S0219477502000981,Khan2018,GUO2008318}), and not a few quantum games induce nontrivial results that are not expected classically.

These ideas of quantizing the strategies can also be applied to gambling theory. A fair quantum gambling game played by two players far from each other was proposed by \cite{PhysRevLett.82.3356}. In the classical case, it is difficult to ensure fair play without introducing a third party to oversee the game. However, by providing a quantum superposition state and then observing it, it is possible to achieve fair gambling. In addition, procedures were devised to protect the security of such remote gambling from entanglement attacks \cite{PhysRevA.64.064302,PhysRevA.66.052311}. Zhang {\it et al.} devised and realized an optical quantum circuit that helps achieve fair gambling and experimentally showed that the error is practically under control \cite{Zhang_2008}. In the sense of quantizing a game, some quantized games have been proposed \cite{Schmidt_2012,SCHMIDT2013400,FRACKIEWICZ20148,qchess,Fuchs_2020,bleiler2009quantized}. Quantum duel was proposed \cite{Schmidt_2012}, and a Russian roulette game using a quantum gun was proposed and studied \cite{SCHMIDT2013400,FRACKIEWICZ20148}. Quantum chess was devised as a more casual game \cite{qchess}. Quantized strategies in poker, one of the most popular card games, were also considered in \cite{bleiler2009quantized} using Meyer's quantized procedure \cite{PhysRevLett.82.1052}. Recently, a game similar to poker was presented on a quantum computer by \cite{Fuchs_2020}. In their experiment, the researchers investigated the expected value under quantum noise on a practical quantum computer. They also showed that the expected value can be improved using quantum error-mitigation techniques. These studies suggest that quite a few games based on quantum mechanics have been examined.

Recent studies have shown that quantum bits yield advantages to blackjack, which is one of the most popular casino games \cite{PhysRevA.102.012425}; the researchers showed that the expectation values of two cooperating persons can increase by imparting information of each individual's card with entangled qubits. In the present paper, strategies and cards are represented as qubits and we devised a quantum circuit for reproducing classical blackjack. Classical blackjack has been studied statistically for more than half a century \cite{doi:10.1080/01621459.1956.10501334,BOND1974413,Thorp1966BeatTD}, and it is decided whether the player should do {\it hit} or {\it stand}. This is called {\it Basic Strategy}. Moreover, it is widely known in various rules, such as number of decks, options, and refunds when the player has blackjack \cite{Thorp1966BeatTD,wizard}. In blackjack, as the player can choose strategies (e.g., hit or stand), we can quantize not only the game itself but also the strategies involved using a method similar to \cite{PhysRevLett.82.1052,PhysRevLett.83.3077}. This quantum circuit reproducing blackjack can be easily realized in the near future when quantum computers are commonplace. Furthermore, we showed that the player's expectation value increases substantially when quantum strategies are employed in the presence of entanglement.

In the next section \ref{rule}, we explain the blackjack rule that we adopted (simplified for ease of calculation). In section \ref{cbj}, we reformulate the blackjack in order to regard it as a object of the quantum game theory. In section \ref{qbj}, we propose a quantum circuit that reproduces blackjack, (especially a toy model,) and express the possibility of entanglement. In section \ref{res}, we show classical expectation value and basic strategy; we also show that using quantum strategies in the presence of entanglement, the player's expectation value increases compared to the classical game. Some conclusions are shown in section \ref{conclusion}.

\section{\label{rule}Blackjack rules and toy model}
Blackjack is a one-on-one card game between a dealer (Alice) and a player (Bob). One deck comprises 52 cards excluding jokers. Aces are worth 1 (hard) or 11 (soft), face cards (JQK) are worth 10, and every other card has its number value.
After the player has placed his bet, the dealer distributes two cards to the player. Similarly, she distributes two cards to herself. At this time the dealer has one card face-up and the other face-down. The objective is to get a hand with a total closer to 21. It is worth noting however, that the 21 hand which is made of one ace and one face card or 10 is the strongest (called natural). 

First, the player must choose to hit or stand. If he chooses to hit, he draws a card. If he exceeds 21 (bust), he loses his bet regardless of the dealer's hand. Next, the dealer opens the face-down card, and if her total is soft 16 or less, she must hit. She repeats this procedure until her total reaches soft 17 or more. If the dealer busts and the player does not, or if the player's hand is closer to 21 than the dealer's, the player wins the bet. If the player's total equals the dealer's total, the bet will be returned (push). In blackjack, the player can do his best by calculating the expected value of the profit that he will obtain in each initial state. This is widely known as a basic strategy and has been calculated in various rules \cite{Thorp1966BeatTD,wizard}. Note that only the player can choose whether to hit or stand. Conversely, the dealer has no choice of a strategy since she just draws the card mechanically until her total reaches soft 17 or more. At first glance, the player who has many choices seems to have a competitive edge. However, the dealer has the advantage because the player loses if both hands exceed 21. The reason many casinos adopt option rules, such as double down, split, and surrender is to increase the player's expectation.

For simplicity, we quantized a toy model of blackjack proposed by Ethier \cite{Ethier2010} and called {\it snackjack} by Epstein \cite{epstein}. As snackjack is highly simplified compared to blackjack which uses 52 cards, it is easy to calculate the basic strategy by hand. Recently, Ethier calculated the expectation and basic strategy under various options, with card counting, and with bet variation in this classical snackjack \cite{ethier2019snackjack}. Snackjack uses eight cards AA223333, aiming toward a maximum value of 7, with an ace having a value of either 1 or 4, and 2 and 3 having their own values. In this paper, we ignored various options and simplified them for calculation. Tab. \ref{tab1} shows the rules we followed.

\begin{table}[h]
    \centering
    \caption{Our rulesets}
    \begin{tabular}{|c||c|} \hline
     ace means & 1 or 4 \\ \hline
     aim to & 7 \\ \hline
     natural is & ace and 3 \\ \hline
     player chooses  & hit or stand (only one time) \\ \hline
     dealer stands  & soft 6 or more \\ \hline
     blackjack pays & 1 to 1  \\ \hline
    \end{tabular}
    \label{tab1}
\end{table}

\section{\label{cbj}Classical snackjack as a quantum game}
Referring to \cite{doi:10.1080/09500340008232180,doi:10.1142/S0219749904000092}, we formulate the blackjack played by a player and a dealer as a {\it 2 player quantum game}.
Blackjack is regarded as the game in which the pay-off is determined by the initial hand cards of the player and the dealer.
More precisely, blackjack is a mapping that assigns a {\it 2 player quantum game} $\Gamma_{p,d}$ to an element $(p,d)$ of $T_{1}\times T_{2}$ where $T_{1}$ is a set of a pair of initial hand cards of the player, $T_{2}$ is a set of an up card of the dealer.
The quantum games $\Gamma_{p,d}=(\mathcal{H},\rho,\Sigma_1,\Sigma_2,P_1,P_2)$ are specified by Hilbert space $\mathcal{H}$ for the states, the initial state $\rho\in\mathcal{H}$, the sets $\Sigma_1$ and $\Sigma_2$ of the strategies of the player and the dealer, and the player's utility $P_{1}$ and the dealer's utility $P_{2}$.
Each player has a qubit and the state of the system is represented as the tensor product of their qubit.
The strategies $\hat{S}_1(\in \Sigma_1)$ and $\hat{S}_2(\in \Sigma_2)$ selected by players are regarded as $2\times 2$ unitary operators acting on their own qubit.
In this formalism, classical blackjack is the quantum game specified by
\begin{gather}
	\mathcal{H}=\left\{a_{00}\ket{0}\otimes\ket{0}+a_{01}\ket{0}\otimes\ket{1}+a_{10}\ket{1}\otimes\ket{0}+a_{11}\ket{1}\otimes\ket{1}|\,a_{00},a_{01},a_{10},a_{11}\in\mathbb{C}\right\}, \notag \\
	\rho = \ket{0}\otimes\ket{1}, ~~~\Sigma_1 = \{\hat{I},\hat{X}\}, ~~~\Sigma_2 = \{\hat{I}\},
\end{gather}
where $\ket{0}$ ($\ket{1}$) denotes stand (hit) and $\hat{X}$ is the first Pauli matrix. The player has the first qubit and the dealer has the second. We will call each of them {\it strategy bit}.
Note that the hit $\ket{1}$ for the dealer means that he draws a card if the score of his hand is less than 6.
Since the utilities are determined by $(p,d)\in T_{1}\times T_{2}$ and player's strategy $\hat{S}_{1}\in\Sigma_{1}$, we can write them as $P_1(p,d,\hat{S}_1)$ and $P_2(p,d,\hat{S}_1)$.
Concretely,
	\begin{align}
	P_1(p,d,\hat{S}_1)
	=E_{\mathrm{std}}\bra{0}\otimes\bra{1}\left(\hat{S}_1\otimes\hat{I}\right)\rho+E_{\mathrm{hit}}\bra{1}\otimes\bra{1}\left(\hat{S}_1\otimes\hat{I}\right)\rho,
	\label{utiladd1}
	\end{align}
where $E_{\mathrm{std}}$ and $E_{\mathrm{hit}}$ are pay-offs corresponding to stand and hit.
It is a point to notice that $E_{\mathrm{std}}$ and $E_{\mathrm{hit}}$ depend on the initial hand cards $(p,d)$.
The utilities do not have $\hat{S}_{2}$ dependence since $\Sigma_{2}=\{\hat{I}\}$ in the classical blackjack. Especially in our snackjack (Tab. \ref{tab1}), $P_2(p,d,\hat{S}_1)=-P_1(p,d,\hat{S}_1)$, thus it is a zero-sum game.

\section{\label{qbj}Entanglement and quantum circuits}
\subsection{\label{entangle}The case of the presence of entangle}
\begin{figure}[t]
     \centering
     \includegraphics[scale=0.8]{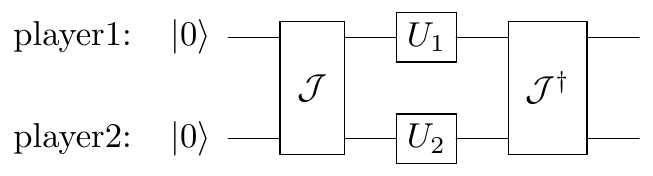}
     \caption{EWL quantization scheme}The gate $\mathcal{J}$ entangles the two qubits and $\mathcal{J}^\dagger$ disentangles those. The players choose an unitary operator $U$ as a quantum strategy.
     \label{add1}
 \end{figure}

In this section, using EWL quantization protocol \cite{PhysRevLett.83.3077}, we investigate the quantum blackjack including an entanglement between the player and the dealer. Eisert {\it et al.} prepared the entangled player's strategies in 2 players symmetric game, especially in PD game (Fig. \ref{add1}). They showed that the dilemma disappears by using restricted quantum strategies and an entanglement. After their work, in PD game, some studies about generalized strategy spaces \cite{PhysRevLett.87.069801,PhysRevLett.88.137902,Du_2003,ICHIKAWA2007531,Ichikawa_2008} or repeated games \cite{IQBAL2002541,Frackiewicz_2012,aoki2020repeated} have been worked. In generalized strategy spaces, the players are not able to escape from the dilemma \cite{PhysRevLett.87.069801,Ichikawa_2008}, and van Enk {\it et al.} \cite{PhysRevA.66.024306} pointed out that EWL protocol does not preserve the non-cooperative aspects of PD. Our blackjack game, however, does not have any dilemmas due to the asymmetry of only the player seeking a win and trying to beat the dealer who has limited strategy space $\Sigma_2=\{\hat{I}\}$. Thus we straightforwardly adopt EWL protocol. In section \ref{conclusion}, we will comment on the degree of freedom of the dealer's strategy in this quantized blackjack game.
 
We consider entanglement between the player's and dealer's strategies. When the player selects the strategy $\hat{S_{p}}$, the state of the system just before the observation is
\begin{equation}
    \mathcal{J}^\dagger (\hat{S}_p \otimes \hat{I})\mathcal{J} \ket{0} \otimes \ket{1},
\end{equation}
where $\mathcal{J}$ is $4\times4$ complex matrix acting on both qubits.

Classically, all strategies the player can choose are $\hat{X}$ or $\hat{I}$. We assume 
\begin{equation}
\mathcal{J} = \mathrm{exp}(-i\frac{\gamma}{2} \hat{X} \otimes \hat{U})
\end{equation}
so that the game reproduces the classical game if the player selects the classical strategies.
The first part is the player's strategy sector and the second is the dealer's, and operator $\hat{U}$ is unitary and hermitian. $\gamma$ is the measure for the game's entanglement \cite{PhysRevLett.83.3077} and we will call $\gamma$ {\it entangle intensity} below. When $\hat{S}_1 = \hat{X} \mspace{5mu} \mathrm{or} \mspace{5mu} \hat{I}$, it follows
\begin{equation}
\mathcal{J}^\dagger ( \hat{X} \otimes \hat{I} ) \mathcal{J} = \hat{X} \otimes \hat{I},
\end{equation}
\begin{equation}
\mathcal{J}^\dagger ( \hat{I} \otimes \hat{I} ) \mathcal{J} = \hat{I} \otimes \hat{I}.
\end{equation}
In this way, without breaking the classical game, we can insert the operator making entangle strategies.

A unitary and hermitian operator is written by
\begin{equation}
\hat{U} = s_0 \hat{I} + s_1 \hat{X} + s_2 \hat{Y} + s_3 \hat{Z} 
\label{Uop1}
\end{equation}
where
\begin{equation}
(s_0^2 + s_1^2 + s_2^2 + s_3^2)\hat{I} +2s_0(s_1\hat{X} + s_2\hat{Y} + s_3\hat{Z}) = \hat{I} \ \  (s_\mu \in \mathbb{R}).
\label{Uop2}
\end{equation}
In this $\Sigma_2=\{\hat{I}\}$ game, except for relative phases, $\hat{I}$, $\hat{Z}$ and $\hat{X}$, $\hat{Y}$ have the same effect on a strategy bit respectively, hence we can set $s_0 = s_2 = 0,s_1 = \sin\theta , s_3 = \cos\theta$ without losing generality. Here $\theta$ is {\it game parameter}.

In the case of general player's strategy $\hat{S}_1$, it follows
\begin{equation}
\begin{split}
\left( \mathcal{J^\dagger}(\hat{S_1} \otimes \hat{I}) \mathcal{J} \right) \ket{0} \otimes \ket{1} = \left\{ \cos^2{\frac{\gamma}{2}} \hat{X}\{\hat{X},\hat{S_1} \} + (\sin^2{\frac{\gamma}{2}} - \cos^2{\frac{\gamma}{2}}) \hat{X}\hat{S_1}\hat{X} \right\} \ket{0} &\otimes \ket{1}  \\
+ i\cos{\frac{\gamma}{2}}\sin{\frac{\gamma}{2}}\sin{\theta} [ \hat{X} , \hat{S_1} ] \ket{0} &\otimes \ket{0} \\ -i\cos{\frac{\gamma}{2}}\sin{\frac{\gamma}{2}}\cos{\theta} [ \hat{X} , \hat{S_1} ] \ket{0} &\otimes \ket{1}.
\end{split}
\end{equation}
From here, we limit $\Sigma_1$ to $\{\hat{I},\hat{X},\hat{Y},\hat{Z}\}$. Both strategies bits after through $\mathcal{J}^\dagger$ are
\begin{align}
\begin{aligned}
\label{iop}
\hat{S_1} &= \hat{I} ~:~\ket{0}\otimes\ket{1} \\
\hat{S_1} &= \hat{X} :~\ket{1}\otimes\ket{1},
\end{aligned}
\end{align}
\begin{align}
\begin{aligned}
\label{zop}
\hat{S_1} &= \hat{Y} :~i\cos\gamma \ket{1} \otimes \ket{1} + \sin\gamma\cos\theta \ket{0} \otimes \ket{1} - \sin\gamma\sin\theta \ket{0} \otimes \ket{0} \\
\hat{S_1} &= \hat{Z} :~\cos\gamma \ket{0} \otimes \ket{1} - i\sin\gamma\cos\theta \ket{1} \otimes \ket{1} + i\sin\gamma\sin\theta \ket{1} \otimes \ket{0}.
\end{aligned}
\end{align}
The state of 01 means stand, and 11 means hit. As it is clear from (\ref{zop}), when the player chooses $\hat{Y}$ or $\hat{Z}$, he can do operations 00 and 10 that do not exist in the classical game.

We summarize the quantized blackjack, employing the formalism of the 2 player quantum game $\Gamma_{p,d}=(\mathcal{H},\rho,\Sigma_1,\Sigma_2,P_1,P_2)$ introduced in the previous subsection.
In order to get the quantized blackjack, we modify two settings of the classical blackjack.
The modifications are expanding the set of player's strategy to $\Sigma_{1}=\{\hat{I},\hat{X},\hat{Y},\hat{Z}\}$ and changing the utilities to
	\begin{align}
	P_1(p,d,\hat{S}_1)
	~=~~&E_{00}\bra{0}\otimes\bra{0}\mathcal{J}\left(\hat{S}_1\otimes\hat{I}\right)\mathcal{J}^{\dagger}\rho
	 +E_{\mathrm{std}}\bra{0}\otimes\bra{1}\mathcal{J}\left(\hat{S}_1\otimes\hat{I}\right)\mathcal{J}^{\dagger}\rho \notag \\
	+~&E_{10}\bra{1}\otimes\bra{0}\mathcal{J}\left(\hat{S}_1\otimes\hat{I}\right)\mathcal{J}^{\dagger}\rho
	 +E_{\mathrm{hit}}\bra{1}\otimes\bra{1}\mathcal{J}\left(\hat{S}_1\otimes\hat{I}\right)\mathcal{J}^\dagger \rho
	 \label{qutility}
	\end{align}
where $E_{00}$ and $E_{10}$ are the expectation values corresponding to operations 00 and 10 respectively. From (\ref{iop})-(\ref{qutility}), $P_1$ corresponding to each strategy $\{\hat{I}, \hat{X}, \hat{Y}, \hat{Z}\}$ is 
\begin{equation}
\label{exp}
\{E_\mathrm{std},E_\mathrm{hit},\sin^2\gamma(\cos^2\theta E_\mathrm{std} +\sin^2\theta E_{00}) +\cos^2\gamma E_\mathrm{hit}, \cos^2\gamma E_\mathrm{std} +\sin^2\gamma(\cos^2\theta E_\mathrm{hit}+\sin^2\theta E_{10}) \}.
\end{equation}

\subsection{\label{q-circuit}Quantum circuit reproducing blackjack}
\begin{figure*}[t]
\centering
\includegraphics[width=15.5cm]{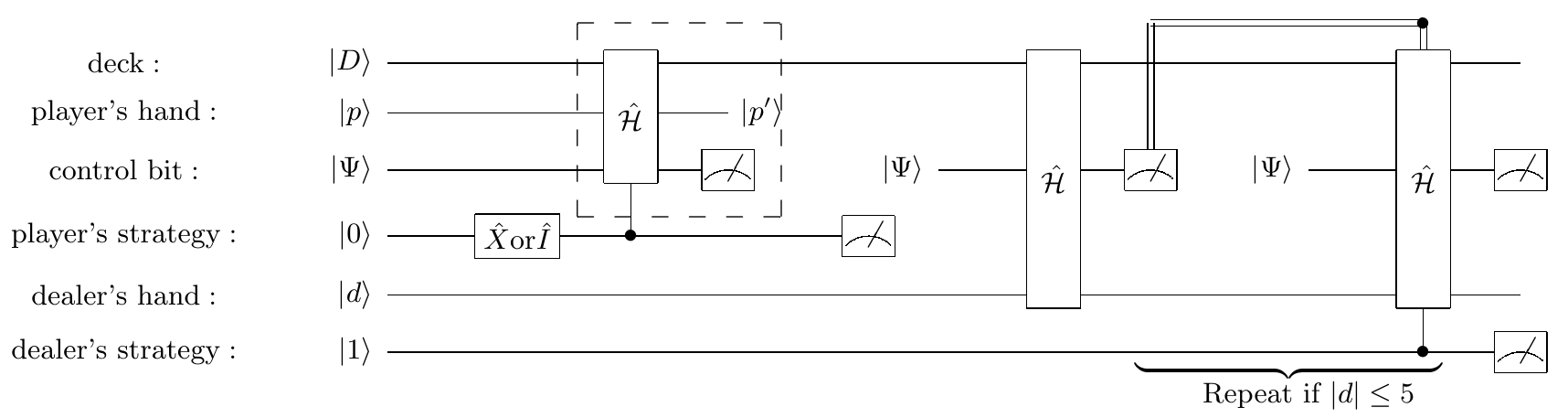}
\caption{Quantum circuit reproducing the classical game}
\label{circuit1}
\end{figure*}

\begin{figure}[t]
\centering
\includegraphics[width=11cm]{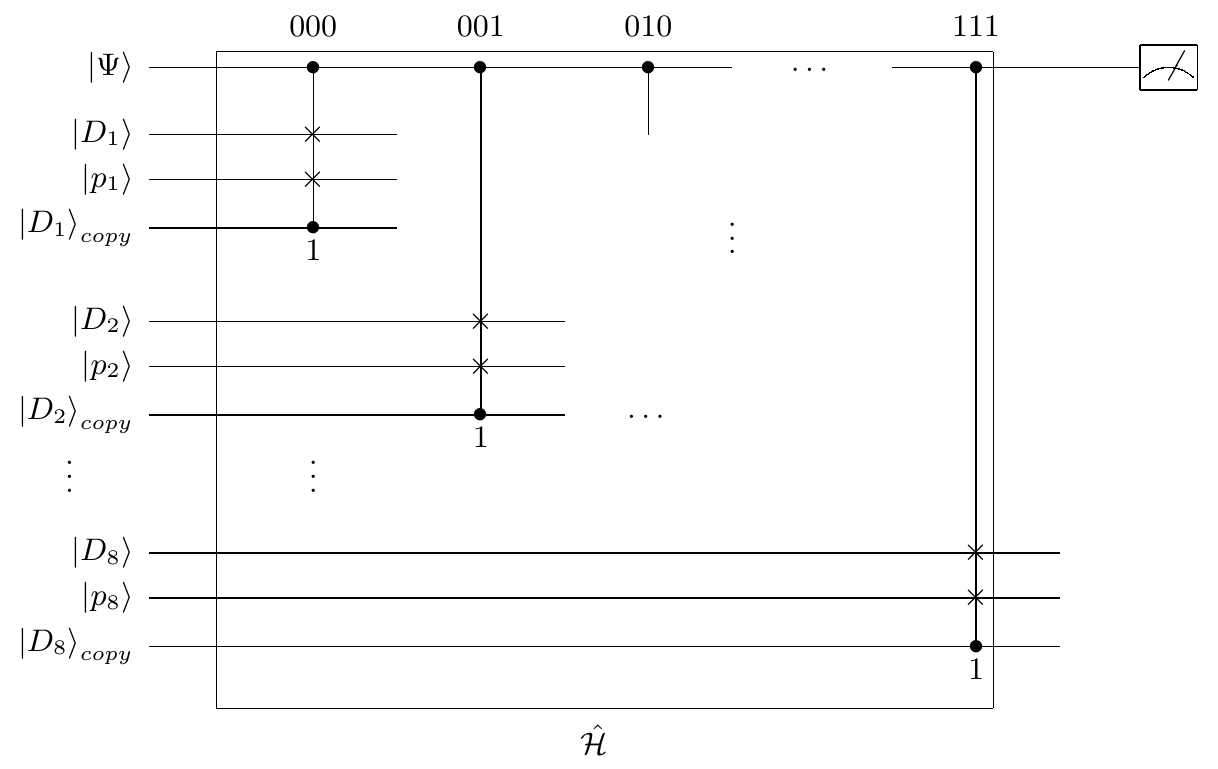}
\caption{Detail of {\it Hit} operator }
\label{circuit2}
\end{figure}

In this section, we show the circuit which reproduces the games discussed above. First of all, we introduce the circuit which reproduces the classical blackjack defined in section \ref{cbj}. Fig. \ref{circuit1} and Fig. \ref{circuit2} show the quantum circuit.
Let $\mathcal{H}_{\rm whole}$ be Hilbert space, on which the circuit in Fig. \ref{circuit1} works and consists of some spaces as follows:
    \begin{align}\label{whole Hilbert}
    \mathcal{H}_{\rm whole}=\mathcal{H}_{\rm deck}\otimes\mathcal{H}_{\rm deck\,copy}\otimes\mathcal{H}_{\rm p\mathchar`-hand}\otimes\mathcal{H}_{\rm control}\otimes\mathcal{H}_{\rm p\mathchar`-strategy}\otimes\mathcal{H}_{\rm d\mathchar`-hand}\otimes\mathcal{H}_{\rm d\mathchar`-strategy}.
    \end{align}
We in turn explain the component spaces of $\mathcal{H}_{\rm whole}$ appearing in eq. \eqref{whole Hilbert}.
We first define a deck state $\ket{D}\in\mathcal{H}_{\rm deck}$, a copied deck state $\ket{D}_{\rm copy}\in\mathcal{H}_{\rm deck\,copy}$, an initial player's hand $\ket{p}\in\mathcal{H}_{\rm p\mathchar`-hand}$, and dealer's hand $\ket{d}\in\mathcal{H}_{\rm d\mathchar`-hand}$ to regard $p\in T_{1}$ and $d\in T_{2}$ as qubits. These take this form
\begin{equation}
\ket{A^{(1)}} \otimes \ket{A^{(2)}} \otimes \ket{2^{(1)}} \otimes \ket{2^{(2)}} \otimes \ket{3^{(1)}} \otimes \ket{3^{(2)}} \otimes \ket{3^{(3)}} \otimes \ket{3^{(4)}},
\label{deck}
\end{equation}
where,
\begin{gather}
\ket{A^{(n)}},\ket{2^{(n)}},\ket{3^{(m)}} = \alpha \ket{0} + \beta \ket{1} \;\;\;\;(n=1,2,~m=1,2,3,4,~ \alpha,\beta \in \mathbb{C}), \\
|\alpha|^2 + |\beta|^2 = 1,
\end{gather}
and then $\ket{1}$ means an existing card corresponding to each card sector and $\ket{0}$ means not. $\mathcal{H}_{\rm deck},\mathcal{H}_{\rm deck\,copy},\mathcal{H}_{\rm p\mathchar`-hand}$ and $\mathcal{H}_{\rm d\mathchar`-hand}$ are spanned by these $2^8$ bases. Secondly, $\mathcal{H}_{\rm p\mathchar`-strategy}\otimes\mathcal{H}_{\rm d\mathchar`-strategy}$ is the same as the space $\mathcal{H}$ in section \ref{cbj} and section \ref{entangle}. Finally, $\ket{\Psi}$ consists of three q-bits and lies on $\mathcal{H}_{\rm control}$, which is a vector space of dimension $2^{3}$. Note that in the actual blackjack there is no object to which the element of $\mathcal{H}_{\rm deck\,copy}$ or $\mathcal{H}_{\rm control}$ corresponds. We, however, need to incorporate these two spaces into our circuit due to some technical reasons. We refer to this in detail below.

We explain the outline of Fig. \ref{circuit1} and Fig. \ref{circuit2} as follows. The player's strategy is expressed by a unitary operator $\hat{S_1}$ acting on its strategy bit. In Fig. \ref{circuit1}, first the dealer and player get the cards classically, then the player chooses the strategy $\hat{S}_1 = \hat{X}$ or $\hat{I}$ which means hit or stand respectively. If the player chooses hit, he has $\hat{X}$ operate to his strategy bit. In that case {\it Hit} operator $\hat{\mathcal{H}}$ acts on the deck bit and his hand bit (in Fig. \ref{circuit2}). This operation signifies drawing a card from the deck. Now, we can expands the player's strategy space $\Sigma_1$ to the set of $2 \times 2$ unitary matrices thus he can overlap his strategy bit. For example, when he chooses the Hadamard gate, the cases of hit and stand overlap with a 1/2 probability. As is clear from (\ref{utiladd1}), this does not affect the maximum value expected by the player. After the player's operation, the dealer turns over her face-down card by $\hat{\mathcal{H}}$ operating, and if the total of her hand ($\equiv |d|$) is 5 or less, she must draw a card one more time. This operation is repeated until $|d| \ge 6$.

We next explain the detail of {\it Hit} operator, surrounded by the dotted line in Fig. \ref{circuit1}. Control bit $\ket{\Psi} (\in \mathcal{H}_{\rm control})$ is responsible for the probability of appearance of cards and it is comprised of three qubits and takes the following form
\begin{eqnarray}
\ket{\Psi} &=& \textstyle\sum\limits_{ijk=0}^1 \frac{1}{\sqrt{8}}\ket{ijk} \\ \nonumber
&=& \frac{1}{\sqrt{8}}  ( \ket{000} + \ket{001}  + \ket{010} + \ket{100} + \ket{110} + \ket{101} + \ket{011} + \ket{111} ).
\end{eqnarray}
Each state corresponds to a sector of the card, e.g., 000 to the first ace, 001 to the second ace, 010 to the first deuce , and so on. Therefore, the probability that every particular card is drawn is equal when the unitary {\it Hit} operator acts on the deck and hand bit. In Fig. \ref{circuit2}, $\ket{\cdot_i} (i=1,2,\cdots,8)$ means the sector of the card in (\ref{deck}), e.g., $\ket{\cdot_1}$ corresponds to the sector of the first ace $\ket{A^{(1)}}$. When $\ket{D_1} = \ket{D_1}_{copy} = \ket{1}$, (which indicates the existence of the first ace,) $\ket{D_1}$ and $\ket{p_1}$, corresponding to $\ket{000}$ which is a component of $\ket{\Psi}$, are swapped. Note that, the deck states can be copied without violating the no-cloning theorem \cite{Wootters1982,DIEKS1982271} because the deck state comprises only $\ket{0}$ and $\ket{1}$, which are orthogonal. After measuring the control bit, it is determined which card has been drawn.

Note that, $\ket{\Psi}$ converges on a state that corresponds to the sector of the card that has already been drawn with a 3/8 probability at the time of first hit even though $\hat{\mathcal{H}}$ operates. For example, when a control bit converges on $\ket{000}$ despite $\ket{D_1} = \ket{D_1}_{copy} = \ket{0}$, there is no exchange. To avoid this problem, we ensure that this procedure is repeated until $\ket{p} \neq \ket{p^\prime}$. Even if a strategy bit has $\ket{0}$ component, we can employ this procedure as we know which sector of the card bit was exchanged, by measuring not the hand bit, but the control bit. This is explained below. For example, in the case of the initial cards states 
\begin{align}
&\ket{p}=\ket{A^{(1)},2^{(1)}} \equiv \ket{1}\ket{0}\ket{1}\ket{0}\ket{0}\ket{0}\ket{0}\ket{0}, \notag \\ 
&\ket{d}=\ket{3^{(1)}}\equiv \ket{0}\ket{0}\ket{0}\ket{0}\ket{1}\ket{0}\ket{0}\ket{0}, \notag\\
&\ket{D}=\ket{A^{(2)},2^{(2)},3^{(2)},3^{(3)},3^{(4)}} \equiv \ket{0}\ket{1}\ket{0}\ket{1}\ket{0}\ket{1}\ket{1}\ket{1},
\end{align}
and the overlapped player's strategy bit with Hadamard operator $\hat{H}$ acting on, the first state is  
\begin{equation}
    \ket{p}\ket{D}\ket{\Psi}\ket{d}\hat{H}\ket{0} = \ket{p}\ket{D}\ket{\Psi}\ket{d}\frac{1}{\sqrt{2}}(\ket{0}+\ket{1}),
\end{equation}
where the last state is the player's strategy bit.
Here, we omitted the dealer's strategy bit, and the copy bits. After thorough $\hat{\mathcal{H}}$, the state is
\begin{equation}
\frac{1}{\sqrt{2}}\left\{\ket{p}\ket{D}\ket{\Psi}\ket{d}\ket{0} + \hat{\mathcal{H}}\left(\ket{p}\ket{D}\ket{\Psi}\right)\ket{d}\ket{1}\right\},
\end{equation}
and after measuring the control bit, it will be
\begin{eqnarray}
    \begin{cases} \ket{p}\ket{D}\ket{d}\ket{0} + \ket{p}\ket{D}\ket{d}\ket{1} &: \mspace{6mu} \ket{\Psi} = \ket{000},  \ket{010} \mathrm{or} \ket{110} \\  \ket{p}\ket{D}\ket{d}\ket{0} + \ket{A^{(1)},A^{(2)},2^{(1)}}\ket{2^{(2)},3^{(2)},3^{(3)},3^{(4)}}\ket{d}\ket{1} &: \mspace{6mu} \ket{\Psi} =  \ket{001} \\ \ket{p}\ket{D}\ket{d}\ket{0} + \ket{A^{(1)},2^{(1)},2^{(2)}}\ket{A^{(2)},3^{(2)},3^{(3)},3^{(4)}}\ket{d}\ket{1} &: \mspace{6mu} \ket{\Psi} = \ket{100} \\ \mspace{30mu} \vdots \\ 
     \ket{p}\ket{D}\ket{d}\ket{0} + \ket{A^{(1)},2^{(1)},3^{(4)}}\ket{A^{(2)},2^{(2)},3^{(2)},3^{(3)}}\ket{d}\ket{1} &: \mspace{6mu} \ket{\Psi} = \ket{111}. \end{cases}
\end{eqnarray}

The overall constant has been dropped. When the control bit converges $\ket{000}$, $\ket{010}$ or $\ket{110}$ in the first line, we repeat this one more time. In this way, we can judge whether we must repeat this procedure without measuring the strategy bit or hand bit. In the following discussions, though we omit specifying this, we always perform this procedure.
\clearpage

\begin{figure}[h]
    \centering
    \includegraphics[scale=0.6]{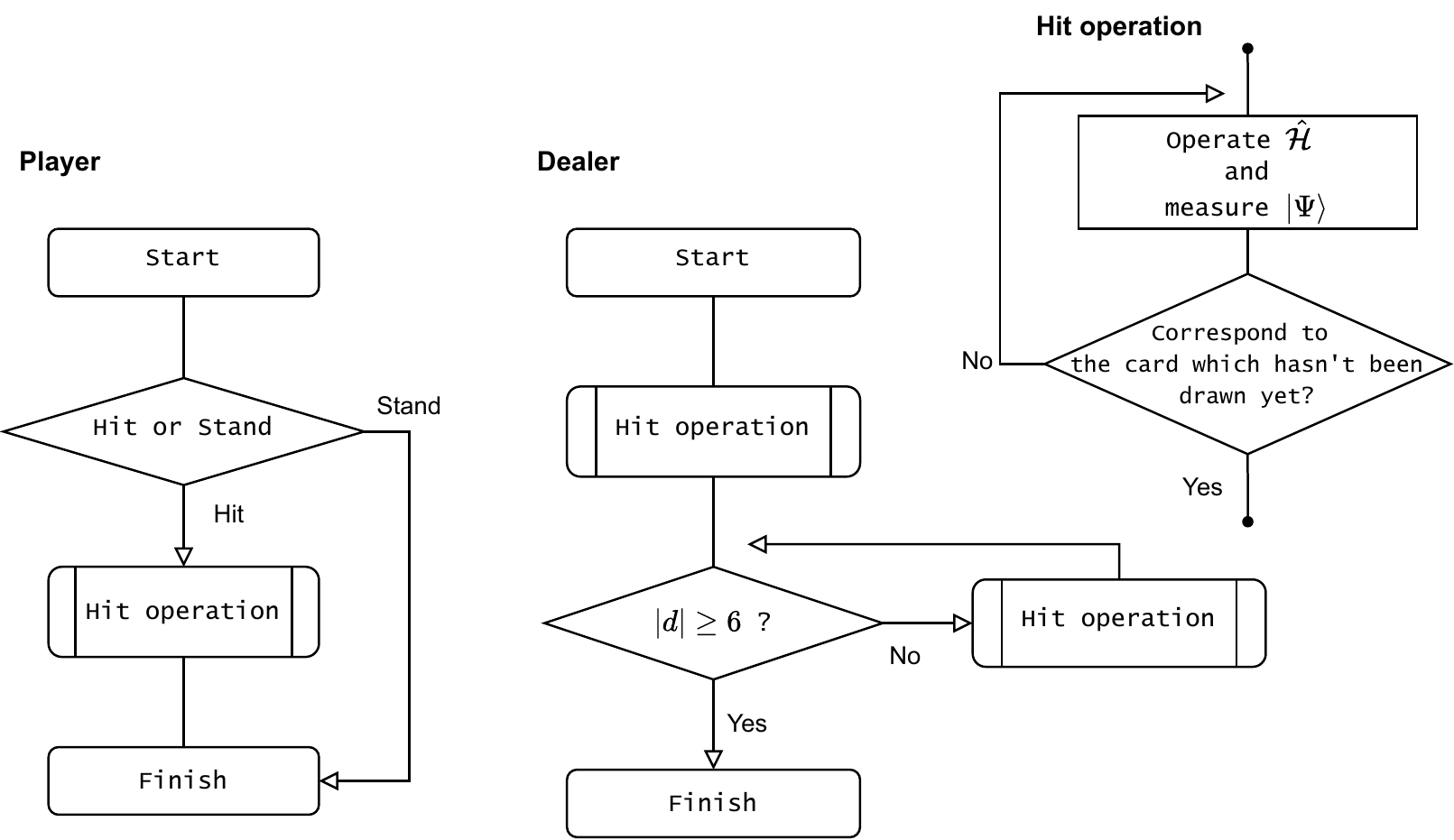}
    \caption{The Process of the game}
    \label{flowchart}
\end{figure}

Dealer's operation {\it Hit} is the same. First, the dealer hits regardless of her strategy bit (corresponding to turn over her face-down card), and if her total $|d|$ is 5 or less and her strategy bit is $\ket{1}$, she repeats the hit operation. Again, we can determine her hand total only by measuring the control bit; therefore, we can build the algorithm where she must hit until her total $|d|$ reaches 6 without measuring her hand bit. Note that when the strategy bit has $\ket{0}$ component, the same algorithm is used. The summary of the algorithm is shown by Fig. \ref{flowchart}.

Finally, after measuring both strategy bits, all states converge. As explained in section \ref{rule}, the player wins when his hand does not exceed 7 and is closer to 7 than the dealer's hand. This quantum circuit reproduces classical snackjack, and therefore the player's expectation and basic strategy must be the same as the classical ones. Such a circuit can be applied to blackjack in a same way by increasing qubits, although it is not considered after this.

\begin{figure}[h]
\centering
\includegraphics[width=15.3cm]{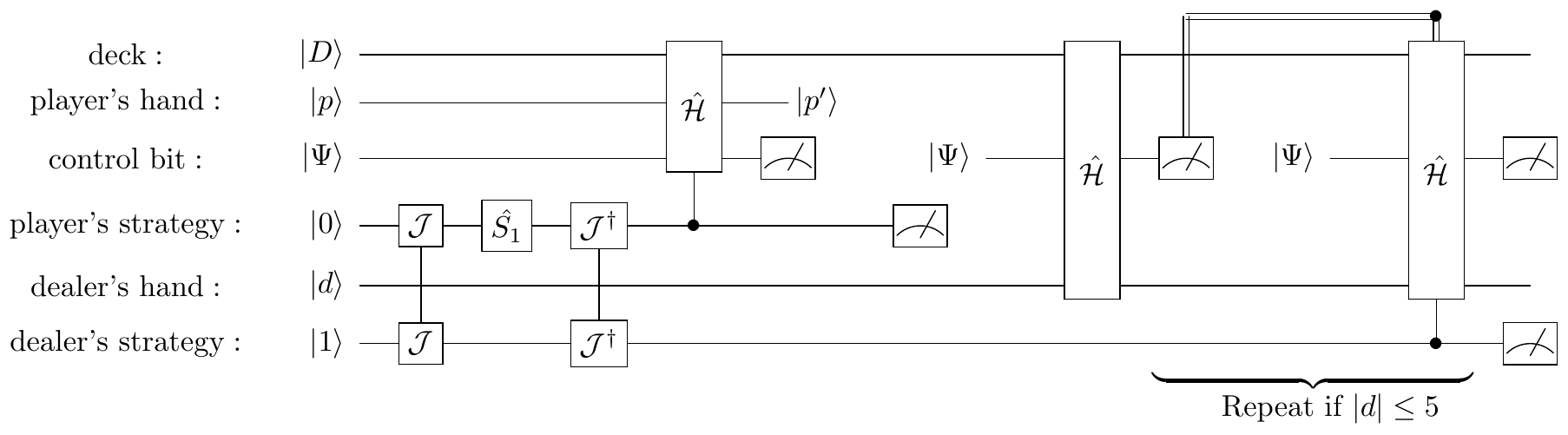}
\caption{Entangled strategy}
\label{fig3}
\end{figure}

We can also design the quantum circuit which reproduces the quantized blackjack formulated in section \ref{entangle}, inserting EWL protocol in  our circuit. The modified quantum circuit is shown in Fig. \ref{fig3}. There is no difference between the circuit in Fig. \ref{fig3} and Fig. \ref{circuit1} except the insertion of EWL protocol.

\clearpage

\section{\label{res}Results}

\begin{figure}[h]
  \begin{minipage}[b]{0.45\linewidth}
    \centering
    \mbox{\raisebox{7.5mm}{\includegraphics[scale=0.6]{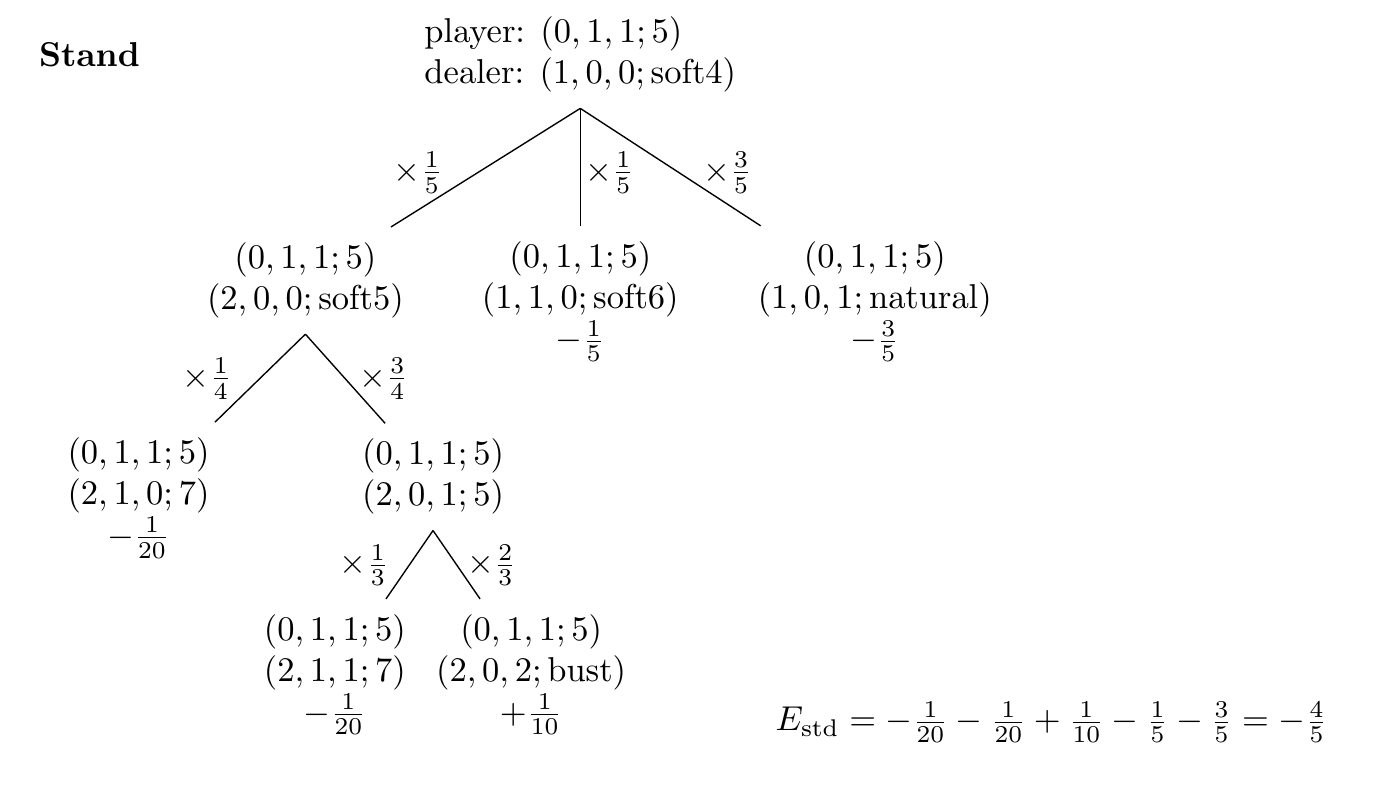}}}
  \end{minipage}
  \begin{minipage}[b]{0.45\linewidth}
    \centering
    \includegraphics[scale=0.6]{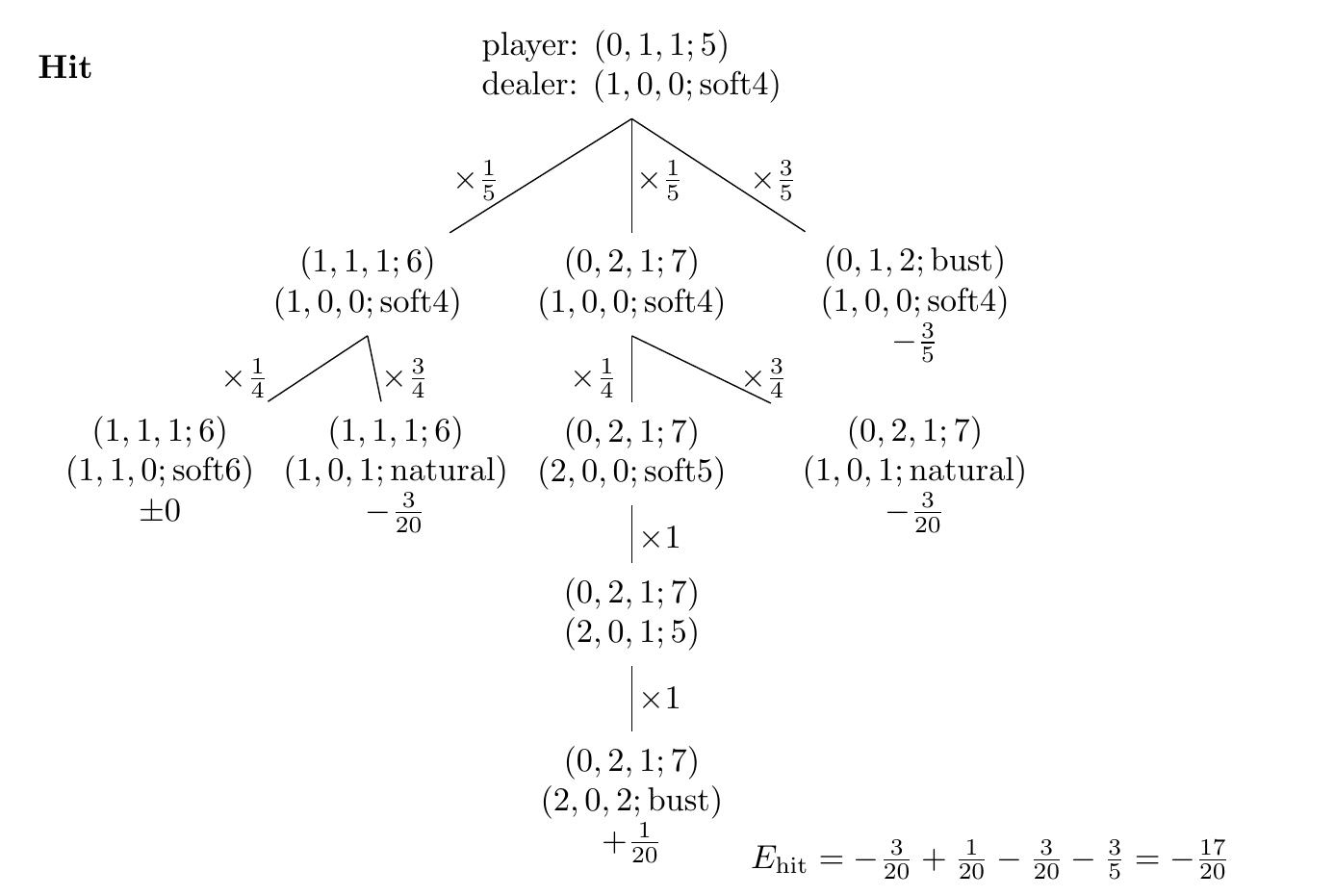}
  \end{minipage}
  \caption{}The calculations of $E_{\mathrm{std}}$ and $E_{\mathrm{hit}}$ in No. 14. The hand states are denoted as (As,2s,3s;total).
  \label{stdandhit}
\end{figure}

Classically, Tab. \ref{tab2} shows $E_{\mathrm{std}},E_{\mathrm{hit}}$ in all initial states and basic strategy. Here, we assume that the player bets unit per one game. Some calculations of $E_{\mathrm{std}}$ and  $E_{\mathrm{hit}}$ in the case of No. 14 game are shown in Fig. \ref{stdandhit}.

\begin{table}[h]
\centering
\scalebox{0.8}[0.8]{
\begin{tabular}{ccccccccc}
No & initial state (As,2s,3s) & dealer up& $E_{\mathrm{std}}$ & $E_{\mathrm{hit}}$ & number of cases & CBS \\ \hline \hline
1 & (2,0,0) & 2 & 1/5 & 1/5 & 2 & I,X \\
2 & & 3 & -2/5 & 1/5 & 4 & X \\
3 & (0,2,0) & 1 & -3/5 & -3/5 & 2 & I,X \\
4 & & 3 & -1 & -2/5 & 4 & X \\
5 &(0,0,2) & 1 & -8/15 & -4/5 & 12 & I \\
6 & & 2 & -1/30 & -1/3 & 12 & I \\
7 & & 3 & -2/5 & -7/15 & 12 & I \\
8 & (1,1,0) & 1 & -4/5 & -4/5 & 4 & I,X \\
9 & & 2 & 3/5 & 3/5 & 4 & I,X \\
10 & & 3 & -1/20 & -1/20 & 16 & I,X \\
11 & (1,0,1) & 1 & 2/5 & -3/10 & 8 & I \\
12 & & 2 & 1 & 1/2 & 16 & I \\
13 & & 3 & 4/5 & 1/30 & 24 & I \\
14 & (0,1,1) & 1 & -4/5 & -17/20 & 16 & I \\
15 & & 2 & -2/5 & -2/5 & 8 & I,X \\
16 & & 3 & -4/5 & -13/30 & 24 & X \\ \hline
total &&&&&168& \\
\end{tabular}
}
\caption{Classical basic strategy (CBS)}No. 1 to 16 are the types of initial states. Each initial state is determined by classical probability.
\label{tab2}
\end{table}

In the classical game, when the player chooses the basic strategy, the expectation value of this game is -1.7\%; thus, the dealer has the advantage in this game. Although the player can also choose $\hat{Y}$ or $\hat{Z}$ in the absence of entanglement, these effects on the strategy bit are the same as $\hat{X}$ and $\hat{I}$, respectively. Therefore, it does not affect the overall expectation value and basic strategy. Note here, that our classical results are different from \cite{ethier2019snackjack}, because of differences in the rulesets (Tab. \ref{tab1}) of snackjack.

\begin{figure}[h]
  \begin{minipage}[b]{0.45\linewidth}
    \centering
    \mbox{\raisebox{7.5mm}{\includegraphics[scale=0.6]{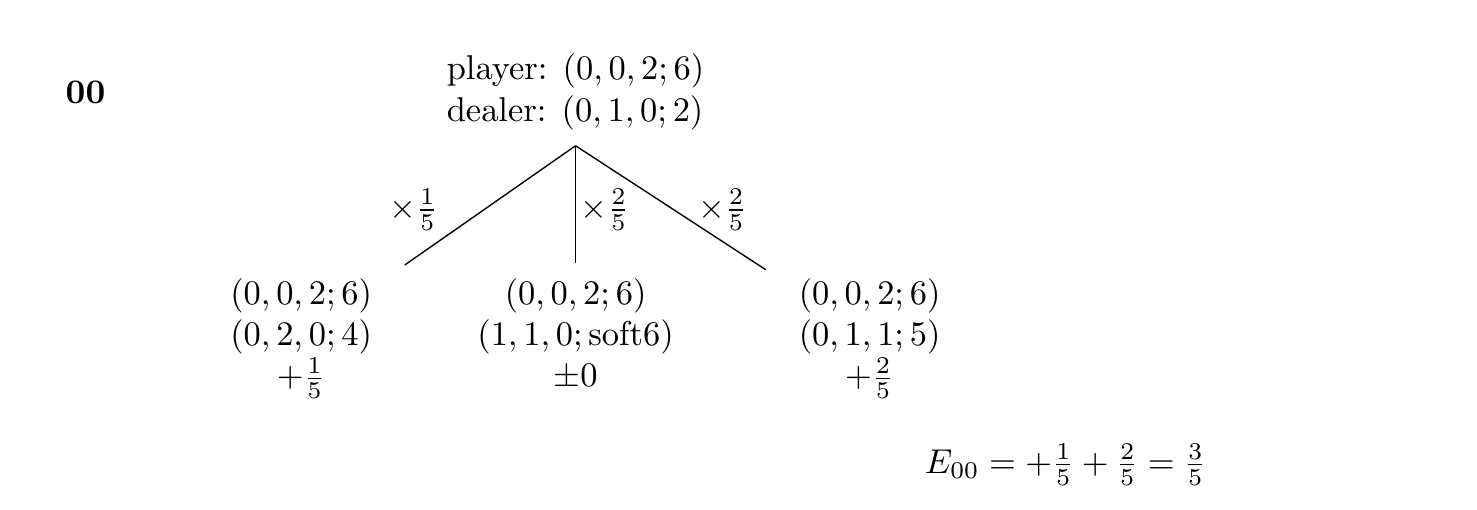}}}
  \end{minipage}
  \begin{minipage}[b]{0.45\linewidth}
    \centering
    \includegraphics[scale=0.6]{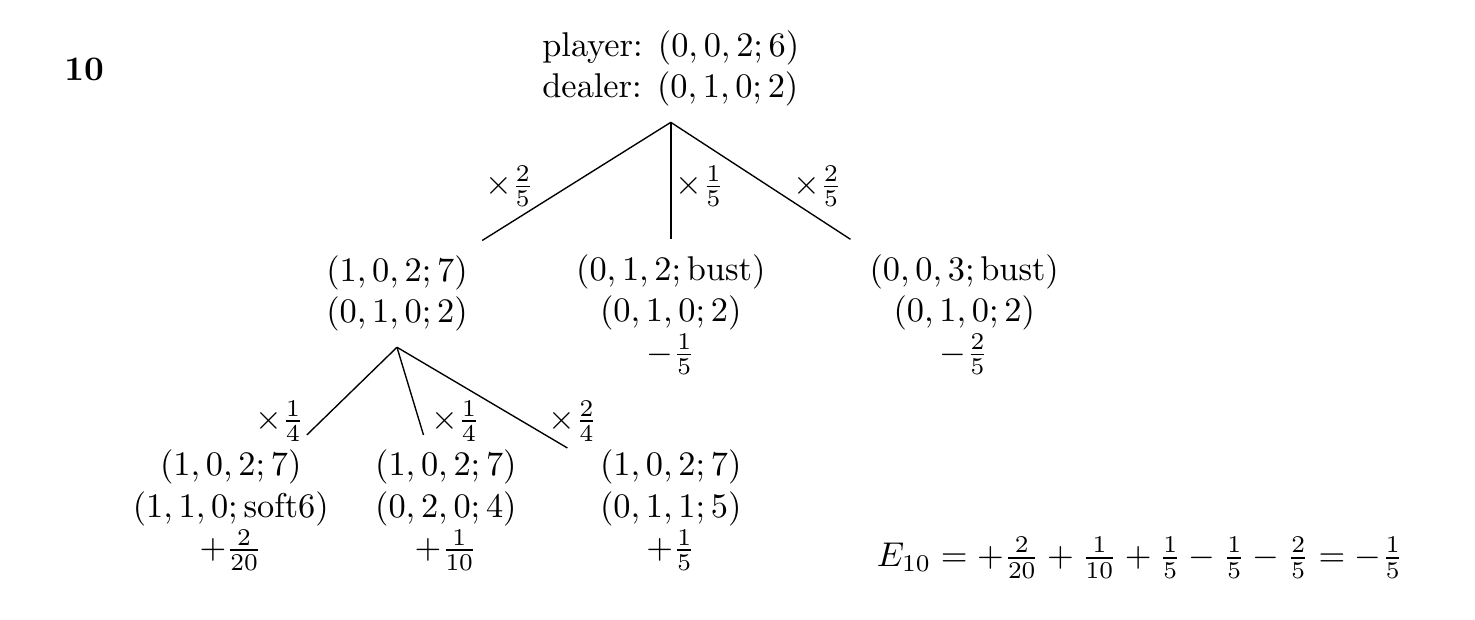}
  \end{minipage}
  \caption{}
  \label{00and10}
\end{figure}

In the presence of entanglement, the player has a meaningful quantum strategy, $\hat{Y}$ or $\hat{Z}$. By the previous section (\ref{zop})-(\ref{exp}), when parameter $\gamma = \theta = \frac{\pi}{2}$, expectation value $E_Y$ which player chooses strategy $\hat{Y}$, equal to $E_{00}$. And also $E_Z = E_{10}$. Tab. \ref{tab3} shows quantum basic strategy (QBS) which the player can choose in entangled strategy. Fig. \ref{00and10} shows the calculations of $E_{00}$ and $E_{10}$ in No. 6 game.
\clearpage
\begin{table}[h]
\centering
\scalebox{0.7}[0.7]{
\begin{tabular}{ccccccccc}
No & initial state (As,2s,3s) & dealer up& $E_{\mathrm{std}}$ & $E_{\mathrm{hit}}$ & $E_{Y} = E_{00}$ & $E_{Z} = E_{10}$ & number of cases &QBS \\ \hline \hline
1 & (2,0,0) & 2 & 1/5 & 1/5 & 1/5 & 2/5 & 2 & Z \\
2 & & 3 & -2/5 & 1/5 & -3/5 & 1/10 & 4 & X \\
3 & (0,2,0) & 1 & -3/5 & -3/5 & -1 & -3/5 & 2 & I,X,Z \\
4 & & 3 & -1 & -2/5 & -1 & -2/5 & 4 & X,Z \\
5 &(0,0,2) & 1 & -8/15 & -4/5 & -1/5 & -4/5 & 12 & Y \\
6 & & 2 & -1/30 & -1/3 & 3/5 & -1/5 & 12 & Y \\
7 & & 3 & -2/5 & -7/15 & 0 & -2/5 & 12 & Y \\
8 & (1,1,0) & 1 & -4/5 & -4/5 & -4/5 & -4/5 & 4 & I,X,Y,Z \\
9 & & 2 & 3/5 & 3/5 & 4/5 & 4/5 & 4 & Y,Z \\
10 & & 3 & -1/20 & -1/20 & 0 & 0 & 16 & Y,Z \\
11 & (1,0,1) & 1 & 2/5 & -3/10 & 2/5 & -3/10 & 8 & I,Y \\
12 & & 2 & 1 & 1/2 & 1 & 4/5 & 16 & I,Y \\
13 & & 3 & 4/5 & 1/30 & 4/5 & 1/10 & 24 & I,Y \\
14 & (0,1,1) & 1 & -4/5 & -17/20 & -4/5 & -17/20 & 16 & I,Y \\
15 & & 2 & -2/5 & -2/5 & -2/5 & -3/10 & 8 & Z \\
16 & & 3 & -4/5 & -13/30 & -4/5 & -2/5 & 24 & Z \\ \hline
total &&&&&&&168& \\
\end{tabular}
}
\caption{Quantum basic strategy ($\gamma = \theta = \pi/2$)}
\label{tab3}
\end{table}

\begin{figure}[h]
\centering
\includegraphics[width=9cm]{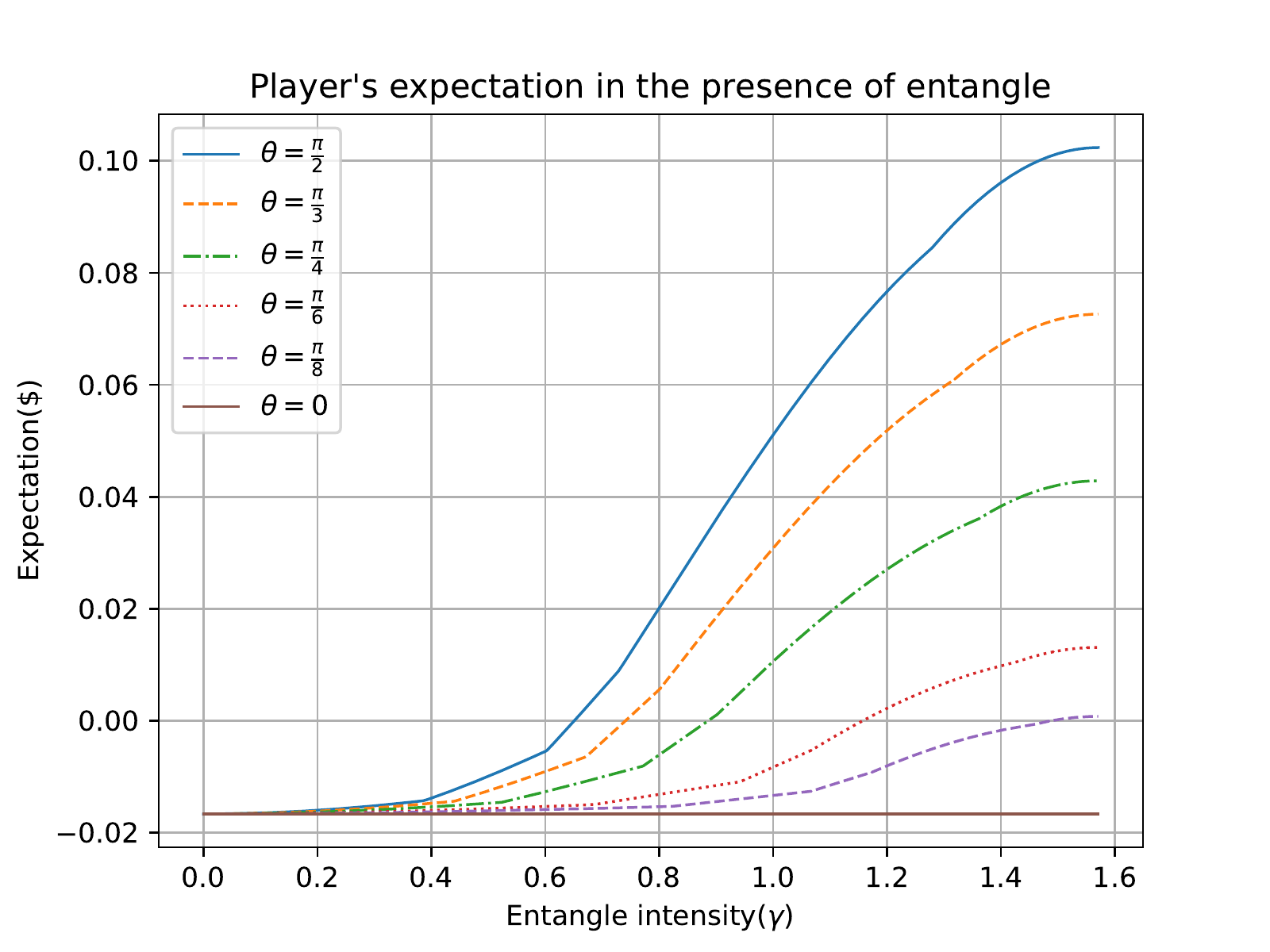}
\caption{Player's expectation in various parameters \\ \centering ($0 \le \gamma , \theta \le \frac{\pi}{2}$)}
\label{fig4}
\end{figure}

When the player chooses QBS in Tab. \ref{tab3}, the overall expected value is +10.2\%; therefore, the player has the advantage in this quantum game. In the $\gamma = \theta = \frac{\pi}{2}$ game, the player has other choices that let the dealer definitely stand (prohibiting the second draw) therefore expectation value increases compared to the classical game. We can set various values in not only entangle intensity $\gamma$ but also game parameter $\theta$. For example in $\gamma = \frac{\pi}{4},\; \theta= \frac{\pi}{2}$, the player's utilities are $\{E_\mathrm{std}, E_\mathrm{hit}, \frac{1}{2}(E_\mathrm{hit}+E_{00}), \frac{1}{2}(E_\mathrm{std}+E_{10})\}$ by (\ref{exp}), and the best choice for the player becomes QBS in each initial hands. The whole expectation for the player in the game is the sum of the largest utility times the appearance ratio of the initial state from No. 1 to No. 16. Fig. \ref{fig4} shows numerical results regarding the player's expectation in various parameters.
As is clear from (\ref{iop})-(\ref{exp}), in $\theta = 0$ game, one can see any $\gamma$ do not change the expectation value from the classical one. Otherwise, when $\gamma = \frac{\pi}{4},\; \theta = \frac{\pi}{2}$, half entangled game, the expectation is +1.8\%. From these results, we conclude that the player has the advantage in the appropriate parameter game.
\clearpage

\section{\label{conclusion}Conclusion and Discussion}
In section \ref{intro}, we introduced quantum effects in game theory \cite{PhysRevLett.83.3077,PhysRevLett.82.1052,doi:10.1142/S0219477502000981,Khan2018,GUO2008318} and the application to gambling \cite{PhysRevLett.82.3356,PhysRevA.64.064302,PhysRevA.66.052311,Zhang_2008,Schmidt_2012,SCHMIDT2013400,FRACKIEWICZ20148,qchess,Fuchs_2020,bleiler2009quantized,PhysRevA.102.012425}. We then proposed a toy model of blackjack, called {\it snackjack} \cite{Ethier2010,ethier2019snackjack,epstein} as a quantum game in section \ref{rule}, \ref{cbj}. In section \ref{qbj}, we found that we can possibly entangle strategies without breaking the classical game and proposed a quantum circuit which reproduced classical blackjack. In section \ref{res}, we showed that the player's expectation increases and the player has an advantage compared to the classical blackjack with QBS.

There remain some interesting questions: (1) The dealer's strategy space $\Sigma_2=\{\hat{I}\}$ can be expanded into more general strategy space, e.g. $\Sigma_2 = \{\hat{I},\hat{Z}\}$. In other words, the dealer inherently has infinite options changing the phase of the initial state $\ket{1}$. This phase affects the player's strategy bit and perhaps cause a dilemma between them depending on the first hand state. In this case, QBS may not be the best choice for the player. (2) It is a nontrivial question that how this quantized blackjack on noisy intermediate scale quantum computers affects the player's expectation, as in their work \cite{Fuchs_2020}. (3) The operation of drawing cards in the circuit we proposed can be applied to other card games. It is also an open question of what kind of quantized casino game will make a difference from the classical one. In the proposed quantum blackjack, we have demonstrated that entanglement between strategies could be used to bankrupt a casino. In the forthcoming quantum era, casinos will have to set house rules that did not exist in the era of classical physics to avoid bankruptcy.

\section*{Acknowledgments}
This study was inspired by a discussion with K. Ikeda and S. Aoki about quantum economics. We acknowledge K. Ikeda for supporting us by proofreading and offering counsel. We also thank T. Onogi for proofreading the manuscript and providing us insightful comments. The authors would like to thank Enago (www.enago.jp) for the English language review.

\bibliography{qbj}% Produces the bibliography via BibTeX.

\end{document}